\documentclass[12pt]{JHEP} 
\usepackage{epsfig}
\newcommand\fverb{\setbox\pippobox=\hbox\bgroup\verb}
\newcommand\fverbdo{\egroup\medskip\noindent%
			\fbox{\unhbox\pippobox}\ }
\newcommand\fverbit{\egroup\item[\fbox{\unhbox\pippobox}]}
\newbox\pippobox

\def\be{\begin{equation}}
\def\ee{\end{equation}}
\def\ba{\begin{array}{l}}
\def\ea{\end{array}}
\def\bea{\begin{eqnarray}}
\def\eea{\end{eqnarray}}

\def\del{\partial}

\def\gap#1{\vspace{#1 ex}}

\title{Bosonization of non-relativstic fermions in $2$-dimensions
and collective field theory}
\author{Avinash Dhar\\
{\it Department of Theoretical Physics,\\
Tata Institute of Fundamental Research,\\ 
Homi Bhabha Road, Mumbai 400 005, India.}
\\
E-mail: \email{adhar@theory.tifr.res.in}}

\preprint{\hepth{0505084}\\
TIFR/TH/05-15}

\abstract{We revisit bosonization of non-relativistic fermions in one
space dimension. Our motivation is the recent work on bubbling
half-BPS geometries by Lin, Lunin and Maldacena
(hep-th/0409174). After reviewing earlier work on exact bosonization
in terms of a noncommutative theory, we derive an action for the
collective field which lives on the droplet boundaries in the
classical limit. Our action is manifestly invariant under
time-dependent reparametrizations of the boundary. We show that, in an
appropriate gauge, the classical collective field equations imply that
each point on the boundary satisfies Hamilton's equations for a
classical particle in the appropriate potential. For the harmonic
oscillator potential, a straightforward quantization of this action can
be carried out {\it exactly} for any boundary profile. For a finite
number of fermions, the quantum collective field theory does not
reproduce the results of the exact noncommutative bosonization, while
the latter are in complete agreement with the results computed
directly in the fermi theory.}


\begin{document}

\section{Introduction} 

The connection between free non-relativistic fermions and string
theory in $2$-dimensions is known since early nineties \footnote{For
an older review of the subject see \cite{GM2}; more recent review and
references can be found in \cite{DKKMMS,TT}.}
\cite{GM1,BKZ,GZJ,GP,SW,MSW,GK,JP,GWM}. Recent studies
\cite{LLM,DB,CS,IM,BCR,AB,NVS,LVW,MM,JT,GMandal,HS,TY,DGOV} have shown
that free non-relativistic fermions also appear in other situations in
string theory, typically in sectors which have high enough
supersymmetry. The requirement of sufficient amount of supersymmetry
is understandable since $2$-dimensional free fermions form an
integrable system. The correspondence with fermions is usually more
easy to see in the dual field theory. Remarkably, in \cite{LLM} the
authors found a class of $1/2-$BPS solutions of supergravity equations
in one-to-one correspondence with classical configurations of free
non-relativistic fermions in $2$-dimensions. While providing yet
another example of AdS/CFT correspondence \footnote{For a review see
\cite{AGMOO}.}, it opens up the interesting possibility of learning
something about the nature of quantum gravity and string theory from
free fermions \cite{GMMPR, GMandal}. Since in \cite{LLM} fermions make
contact with geometry via the bosonized theory which describes their
collective motion, it is essential to understand all aspects of the
bosonized theory in order to be able to use the full power of free
fermions. This provides the main motivation for the present work.

The organization of this paper is as follows. In Section 2 we review
(see also \cite{MW}) the works
\cite{DMW-classical,DMW-nonrel,DMW-path} in which an exact
bosonization of free non-relativistic fermions in $2$-dimensions has
been developed in terms of a noncommutative theory using Wigner phase
space density. In Section 3 we discuss the classical limit of this
bosonized theory. In the classical limit, a generic configuration
consists of droplets of fermi fluid on the phase plane. The dynamics,
which is associated to the collective motion of the droplets,
manifests itself in their changing boundaries. The action for this
collective motion has a built in symmetry under arbitrary
time-dependent reparametrizations of the droplet boundaries. This
gauge symmetry reflects the fact that the motion of the fluid along
the boundaries of the droplets is unphysical because of the
indistinguishability of fermions. In this section, we derive a
manifestly gauge-invariant classical action which describes the
boundary dynamics of the droplets for arbitrary boundary
profiles. Issues related to gauge fixing for different droplet
boundary profiles are also discussed in this section. In an
appropriate gauge, we show that classically each point on the boundary
simply follows Hamilton's equations of motion. Quantization of the
collective field theory is carried out in Section 4. For the harmonic
oscillator potential, quantization can be carried out exactly for any
droplet boundary profile. We find that for a finite number of
fermions, the spectrum of the collective quantized theory does not
agree with the exact spectrum at large energies. Furthermore, the
phase space density fails to reproduce the precise details of the
exact result. We summarize our results and end with some concluding
remarks in Section 5.
 
\section{Review of exact bosonization} 

Consider $N$ free non-relativistic fermions moving in one space
dimension in a potential. The fermion wavefunctions satisfy the
Schroedinger equation
\bea i\hbar~\del_t \psi(x, t) = H \psi(x, t),
\label{twoone}
\eea 
where the single-particle Hamiltonian $H$ is given by 
\bea
H={1 \over 2}(-{\hbar}^2 \del_x^2+V(x)). 
\label{twotwo}
\eea
Moreover, the fixed fermion number constraint is,
\bea
N=\int^{+\infty}_{-\infty} dx~\sum_{m=1}^N 
\psi_m^*(x, t)\psi_m(x, t).
\label{twothree}
\eea 
Here $\psi_m(x, t)~(m=1,2,3 \cdots)$ form a complete orthonormal set of
single-particle wavefunctions,
i.e. $\delta_{mn}=\int^{+\infty}_{-\infty} dx~\psi_m^*(x, t)\psi_n(x,
t)$.

\subsection{Bosonization in terms of Wigner density}

The bosonization carried out in
\cite{DMW-classical,DMW-nonrel,DMW-path} uses the Wigner phase space
density as its basic building block. In terms of the fermion
wavefunctions, it is defined by the expression
\bea u(p, q, t)=\int^{+\infty}_{-\infty} dx~e^{-ipx/\hbar}~
\sum_{m=1}^N \psi_m^*(q-x/2, t)\psi_m(q+x/2, t).
\label{twofour}
\eea

A fundamental property of the Wigner density, one that captures the
underlying fermionic nature of the degrees of freedom in the bosonized
version, is that it satisfies a nonlinear constraint. This constraint,
which can be elegantly written using the non-commutative star
product \footnote{The compact star product notation was not used in
references \cite{DMW-classical,DMW-nonrel,DMW-path}. The expressions
given there are however the same with star products written out in
long-hand.} in the phase plane, is
\bea
u*u(p, q, t)=u(p, q, t),
\label{twofive}
\eea
where the star product is defined in the usual way,
\bea
u_1*u_2(p, q, t)=\biggl[e^{\frac{i\hbar}{2}
(\del_{q_1}\del_{p_2}-\del_{q_2}\del_{p_1})} u_1(p_1, q_1,
t)u_2(p_2, q_2, t)\biggr]_{q_1=q_2=q,~p_1=p_2=p}.
\label{twosix}
\eea 
A quick way of deriving the constraint is to first construct the
bilocal fermion bilinear $\sum_{m=1}^N \psi_m^*(x, t)\psi_m(y, t)
\equiv \phi(x, y, t)$. As a consequence of the orthonormality of the
fermion wavefunctions, this bilocal function satisfies the constraint
$\int^{+\infty}_{-\infty} dz~\phi(x, z, t)\phi(z, y, t)=\phi(x, y,
t)$. The definition of $u$, eqn(\ref{twofour}), which can be easily
rewritten in terms of the bilocal function $\phi(x, y, t)$, and some
simple algebraic manipulations then lead to the constraint
(\ref{twofive}) on $u$.

In addition to the above constraint, $u$ satisfies the condition
\bea
N=\int {dpdq \over 2\pi\hbar}~u(p, q, t),
\label{twoseven}
\eea 
which is just a restatement, in terms of the Wigner density $u$, of the
fact that the total number of fermions is $N$. Finally, an equation of
motion can be derived for $u$ using the Schroedinger equation
satisfied by the fermion wavefunctions. One gets,
\bea
\del_t u(p, q, t) = \{h, u\}_*(p, q, t).
\label{twoeight}
\eea 
Here $h={1 \over 2}(p^2+V(q))$ is the classical single-particle
Hamiltonian. The bracket on the right-hand side is Moyal's
generalization of the Poisson bracket involving the star product,
namely $\{h, u\}_*={1 \over i\hbar}(h*u-u*h)$. In the limit $\hbar
\rightarrow 0$ the Moyal bracket goes over to the Poisson bracket.

\subsection{The role of W$_\infty$ symmetry}

Notice that the constraints (\ref{twofive}) and (\ref{twoseven}) are
left unchanged by the following infinitesimal variation of $u$,
\bea
\delta u=\{\epsilon, u\}_*,
\label{twonine}
\eea
where $\epsilon=\epsilon(p, q, t)$. This is, in fact, the most general
variation that leaves these two constraints unchanged. If the
t-dependence of $\epsilon$ is such that $\del_t \epsilon=\{h, \epsilon
\}_*$~, then $u+\delta u$ also satisfies the equation of motion
(\ref{twoeight}). It is easy to see that two such successive
transformations satisfy the group composition law. In fact, the
relevant group is W$_\infty$, the quantum generalization of the group
of area-preserving diffeomorphisms in $2$-dimensions, 
and these transformations move us on a co-adjoint orbit of W$_\infty$ 
in the configuration space of $u$'s. In particular, changes in $u$
corresponding to t-independent $\epsilon$'s satisfying $\{h, \epsilon
\}_*=0$ are symmetries. The corresponding conserved charges
\footnote{Similar charges first appeared in \cite{SW} in a system of 
free fermions in inverted harmonic oscillator potential, which is  
relevant for the $c=1$ matrix model.} are
\bea
Q_n=\int {dpdq \over 2\pi\hbar} \quad h^n*u(p, q, t), 
\quad n=0, 1, 2, \cdots, 
\label{twoten}
\eea
where $h^n=h*h* \cdots *h$ has $n$ factors. 

The charges $Q_0$ and $Q_1$ are familiar. They measure the total
number of fermions and total energy, respectively. Other charges are
less familiar in the bosonic version. In terms of fermions, however,
they can be easily seen to be sums of higher (than linear) powers of
individual fermion energies, which are obviously conserved in the
non-interacting theory that we are considering here \footnote{In the
fermionic theory, all the charges in any given state can be explicitly
written in terms of the individual energies of $N$ fermions. One might
wonder how the bosonic version of these charges in (\ref{twoten})
depends on only $N$ independent parameters. The point is that the $u$'s
that should be used to evaluate these charges must satisfy the
constraints (\ref{twofive}) and (\ref{twoseven}). Once this is
ensured, it can be seen that the charges $Q_n$ can be expressed in
terms of only $N$ independent parameters.}.

\subsection{The action of the bosonized theory}

Equations (\ref{twofour})-(\ref{twoeight}) constitute an exact
bosonization of the fermion problem. In
\cite{DMW-classical,DMW-nonrel,DMW-path} a derivation of
this bosonization has been given and an action obtained for the Wigner
density, whose variation gives rise to the equation of motion
(\ref{twoeight}). The derivation uses coherent states and the
coadjoint orbits of W$_\infty$, much like the coherent states and
coadjoint orbits of SU$(2)$ are used in the case of a spin in a
magnetic field. As in the latter case, one needs to construct a
``cap'' action, the cap being parametrized by time $t$ and an
additional variable $s$ $(0 \leq s \leq 1)$, such that $u(p, q,
t)=u(p, q, t; s=1)$ while $u(p, q, t; s=0)=\bar u(p, q)$ is a
t-independent function. One gets the following action:
\bea
S=\int dt \int {dp dq \over 2\pi \hbar} \biggl(\int^1_0 ds~\hbar^2 
u*\{\del_t u, \del_s u\}_* - u*h \biggr)
\label{twoeleven}
\eea
It can be easily verified that the equation of motion (\ref{twoeight})
follows from this action upon using the variation (\ref{twonine}) and
the constraints (\ref{twofive}) and (\ref{twoseven}).

A simpler form for the action can be obtained, one that is useful for
going over to the classical limit, if one rewrites the action in terms
of a ``reference'' density $u_0$ (for example, it could be the density
in the fermi vacuum) and the W$_\infty$ group element $v$ 
\footnote{$v$ is unitary, that is $v*v^\dagger=v^\dagger*v=I$.}
that is needed to ``rotate'' $u_0$ to $u=v*u_0*v^\dagger$. One gets
\bea
S=\int dt \int {dp dq \over 2\pi \hbar} \quad u*(a_t - h),
\label{twotwelve}
\eea
where $a_t \equiv i\hbar~\del_t v*v^\dagger$. Since $a_t$ also satisfies
the equation
\bea
\del_t u=\{a_t, u\}_*,
\label{twothirteen}
\eea
which does not depend on the reference density $u_0$, we may alternatively
use this equation to define $a_t$. We will later use the equations 
(\ref{twotwelve}) and (\ref{twothirteen}) to take the classical limit of this
bosonized theory.

\subsection{Consistency of bosonization}

We will end this section by arguing that the equations
(\ref{twofive})-(\ref{twoeight}) provide an exact bosonization in the
sense that solutions to these equations are in one-to-one
correspondence with the states of the fermion system. Consider first
the constraint equation (\ref{twofive}).  General solutions to this
equation have been obtained in \cite{GMS}.  For the harmonic
oscillator potential for which $h=\frac{1}{2} (p^2+q^2) \equiv \rho$,
which is the case of interest in \cite{LLM}, the analysis is
particularly simple. This is because in this case t-independent
solutions of (\ref{twoeight}), which correspond to energy eigenstates,
are radial functions in the phase plane. As discussed in \cite{GMS},
there are an infinite number of real independent radial solutions to
the equation (\ref{twofive}). The general form of the solution is
\bea
u=\sum_{m=0}^\infty c_m \phi_m(\rho), \nonumber 
\eea
where $\phi_m(\rho)=2(-1)^m e^{-2h}L_m(4\rho)$, $L_m$ being the $m$th
Laguerre polynomial. The coefficients $c_m$ can take the values $0,
1$. The fermion number constraint (\ref{twoseven}) fixes the number of
nonzero $c_m$'s to be precisely $N$. These, in fact, identify the
filled levels, so it is clear that solutions to the equations
(\ref{twofive})-(\ref{twoeight}) are in one-to-one correspondence with
the energy eigenstates of the fermion theory. In fact, given a
time-independent $u(p, q)$ which solves the equations
(\ref{twofive})-(\ref{twoeight}), one can uniquely reconstruct the
filled levels. Since $u$ is essentially a projection operator, finding
the corresponding fermi state amounts to finding the subspace on which 
it projects:
\bea
\int {dp dq \over 2\pi \hbar}~e^{ip(x-q)/\hbar}~
u(p, \frac{x+q}{2})~\psi(q) = \psi(x).
\label{twofourteen}
\eea
For $N$ fermions, this subspace is $N$-dimensional, so equation
(\ref{twofourteen}) should have $N$ linearly independent solutions. In
this way one can obtain the wavefunctions for the filled states from
the given $u$.

We give below the solution for the Wigner density in the fermi vacuum:
\bea
u_F=\sum_{m=0}^{N-1} 2(-1)^m e^{-2\rho/\hbar}L_m(4\rho).
\label{twofifteen}
\eea
The expression on the right hand side can be rewritten
\footnote{Actually the answer depends on whether $N$ is even or
odd. The density falls off to zero at infinity only for $N$ odd. It is
this answer that we have given in the equation below.} as an integral
over a single Laguerre polynomial:
\bea
u_F=\int^\infty_{4\rho/\hbar} dy~e^{-y/2} L^1_{N-1}(y).
\label{twosixteen}
\eea
Note that for large $N$ \cite{GR}, 
\bea
e^{-y/2} L^1_{N-1}(y)=\pi^{-1/2}y^{-3/4}N^{1/4}~
{\rm cos}\biggl( 2\sqrt{(N-1)y}-3\pi/4 \biggr)+{\rm O}(N^{-1/4}).
\label{twoseventeen}
\eea
We will make use of these expressions when we compare $u_F$ given above
with the one obtained by quantizing the classical limit of the action
in (\ref{twothirteen}).

\section{Classical limit and collective action}

The easiest way to take the classical limit of the bosonized theory
obtained as outlined above is through the equations (\ref{twotwelve})
and (\ref{twothirteen}), supplemented by the constraints
(\ref{twofive}) and (\ref{twoseven}). To be precise, in the classical
limit we will set $\hbar \rightarrow 0$ and $N \rightarrow \infty$
while keeping $\hbar N = \rho_0$ fixed. In this limit equation
(\ref{twofive}) becomes $u^2=u$, whose standard solutions are filled
and empty regions of phase plane. Equation (\ref{twoseven}) fixes the
total area of the filled regions, $\int dp dq~u = 2\pi \rho_0$. Note
also that the classical energy of any configuration diverges in this
limit as O($1/\hbar$).

\subsection{Classical motion is area-preserving diffeomorphisms}

Let us now first consider equation (\ref{twothirteen}). In the
classical limit, the Moyal bracket on the right-hand side reduces to
the Poisson bracket. Integrating this equation over an infinitesimal
amount of time $\delta t$ gives 
\bea 
u(p-\delta t~\del_q a_t, q+\delta t~\del_p a_t, t+\delta t) 
\approx u(p, q, t). \nonumber 
\eea
This equation shows that the classical motion of the fermi fluid is
determined by area preserving diffeomorphisms in phase space. Here the
diffeomorphism is generated by the function $a_t$. Physical motion of
the fermi fluid is manifested only in changing boundaries
\footnote{This is because in the interior of the filled regions, the
motion is unphysical since the fermions filling the interior are 
identical. Even on the boundary of a filled
region, motion of the fluid along the boundary is unphysical for the
same reason.} of the filled regions. Since the point $(p, q)$ in the
phase plane moves to the point $(p-\delta t~\del_q a_t, q+\delta
t~\del_p a_t)$ in an infinitesimal time $\delta t$, for consistency 
we must have
\bea
\del_t p = -\del_q a_t, \quad \quad \del_t q = \del_p a_t.
\label{threeone}
\eea
We will use these equations below to get an explicit form for the
classical action involving only the boundary variables.

\subsection{The boundary action}

Consider the action in (\ref{twotwelve}) in the classical limit. For
simplicity we will assume that the density $u$ is nonzero only in a
single connected region $\Sigma$, centered at the origin of the phase
plane, as shown in Fig.1(a). Generalization to several disconnected
filled regions, as for example in Fig.1(b), is straightforward.
\begin{figure}[ht]
\begin{tabular}{ll}
{
    \epsfxsize=5.5cm
   \epsfysize=5cm
   \epsffile{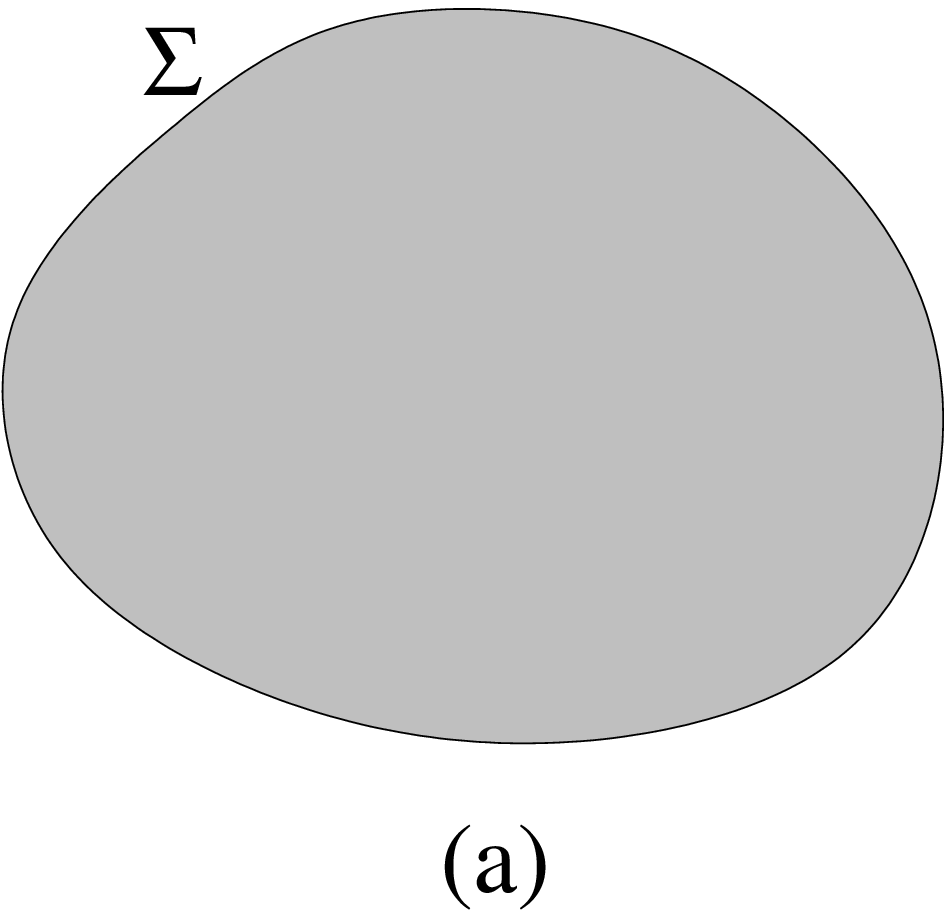}
}
&
{\hspace{2.3cm}
    \epsfxsize=5.5cm
   \epsfysize=5cm
   \epsffile{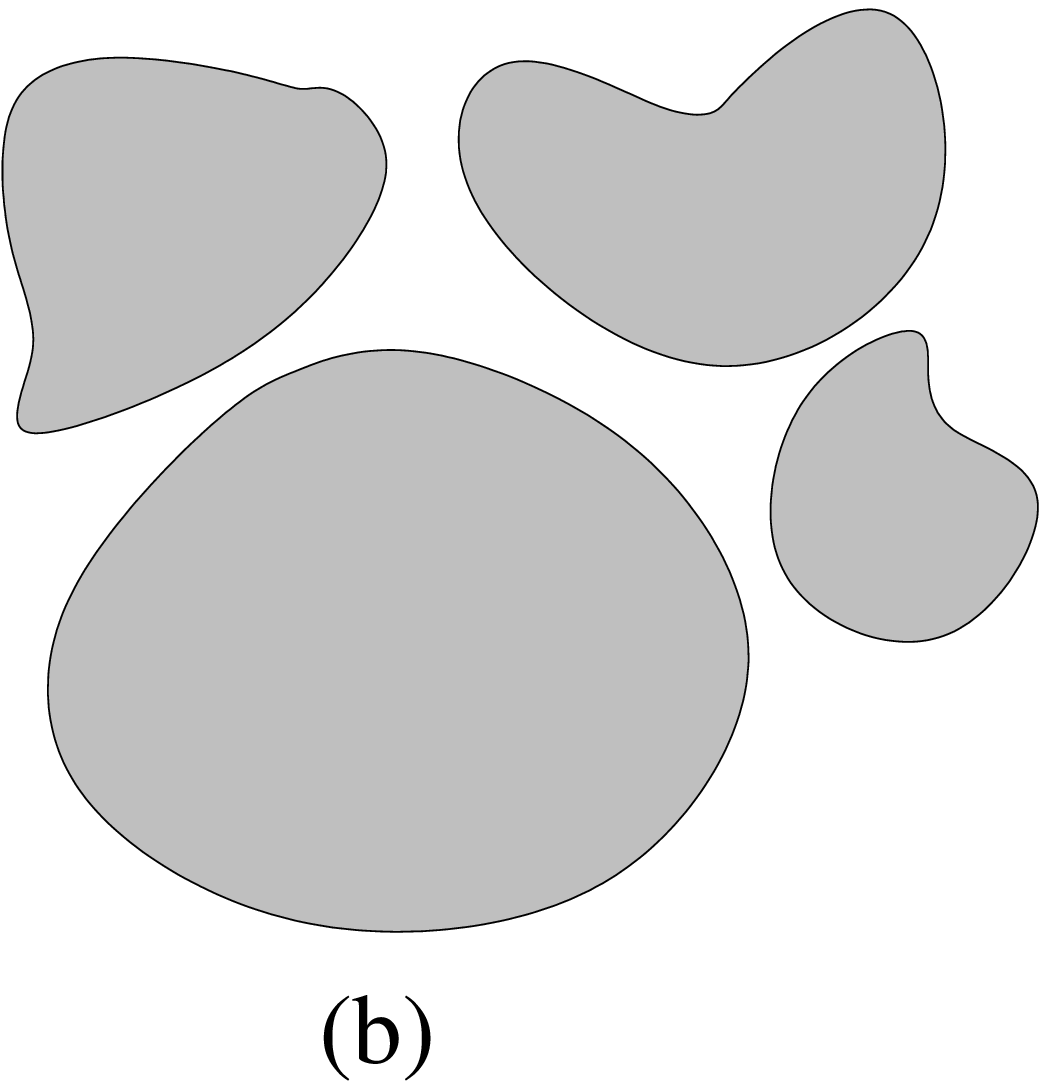}
 }
\end{tabular}
\caption{\sl 
(a) A simple configuration with a single connected filled region
$\Sigma$ in phase space. The droplet is centered at the origin in the
phase plane.  (b) A general configuration of several disconnected
fluid droplets on the phase plane.}
\label{fig1ab.fig}
\end{figure}
Let us parametrize the filled region $\Sigma$ by $(\tau, \sigma)$
which take values in the unit disc $0 \leq \tau \leq 1, \quad 0 \leq
\sigma \leq 2\pi$. Then the classical action can be written as an
action on this disc
\bea 
S=\frac{1}{2\pi \hbar} \int dt \int_0^1 d\tau \int_0^{2\pi} d\sigma
\quad (\del_\sigma p \del_\tau q - \del_\sigma q \del_\tau p) (a_t - h).
\label{threetwo}
\eea
For later purposes, it will be more convenient for us to work in polar 
coordinates \footnote{For applications to $2$-dimensional string theory 
and the $c=1$ matrix model, it would be more appropriate to work in
hyperbolic coordinates.}, $q=r {\rm cos}\theta,~p=r {\rm sin}\theta$. 
In these coordinates the above action becomes
\bea 
S=\frac{1}{2\pi \hbar} \int dt \int_0^1 d\tau \int_0^{2\pi} d\sigma
\quad (\del_\sigma \theta \del_\tau \rho - 
\del_\sigma \rho \del_\tau \theta) (a_t - h),
\label{threethree}
\eea 
where we have used $r^2/2 \equiv \rho$. We also write down the 
consistency conditions (\ref{threeone}) in these coordinates:
\bea
\del_t \theta = -\del_{\rho} a_t, \quad \quad \del_t \rho = 
\del_{\theta} a_t.
\label{threefour}
\eea

Since physical motion of the fermi fluid is manifested only in
changing boundary of the filled region $\Sigma$, it should be possible
to reexpress this classical action in terms of appropriate degrees of
freedom which live only on the boundary of the disc. To do so, let us
introduce the collective field \footnote{The collective field approach
to approximate bosonization of non-relativistic fermions was first
used in \cite{JS}.} $\phi$, which is defined by the relation
\bea 
\rho \del_\sigma \theta = \del_\sigma \phi.
\label{threefive}
\eea
In terms of this variable, 
\bea
(\del_\sigma \theta \del_\tau \rho -
\del_\sigma \rho \del_\tau \theta) = \del_\sigma(\del_\tau \phi - \rho
\del_\tau \theta).
\nonumber
\eea
Also, using the consistency conditions (\ref{threefour}), we get 
\bea
\del_\sigma a_t & = &  \del_t \rho \del_\sigma \theta - 
\del_t \theta \del_\sigma \rho \nonumber \\
& = & \del_\sigma(\del_t \phi - \rho \del_t \theta).
\nonumber 
\eea
With the help of these two relations, and after some algebraic
manipulations involving partial integrations, the integrands of both
the kinetic term involving $a_t$ as well as that of the hamiltonian
piece in action (\ref{threethree}) can be expressed as total
derivatives in $\tau$. The resulting boundary action (at $\tau=1$) is 
\bea
S=\frac{1}{2\pi \hbar} \int dt \int_0^{2\pi} d\sigma
\biggl [\frac{1}{2}\del_\sigma \phi (\del_t \phi - \rho \del_t \theta) -
\del_\sigma \theta \tilde h \biggr],
\label{threesix}
\eea
where $\del_\rho \tilde h=h$ \footnote{For example, for the harmonic
oscillator potential, $h=\frac{1}{2}(p^2+q^2)=\rho$ and so $\tilde
h=\rho^2/2$, upto an irrelevant constant.}. This action, together with
the fixed area constraint, $\int d\sigma~\del_\sigma \theta \rho=2\pi
\rho_0$, describes the dynamics of the boundary of the filled region
$\Sigma$.

\subsection{Boundary reparametrizations}

Remarkably, the equations of motion arising from the above action for
independent variations with respect to $\phi$ and $\theta$ turn out to
be {\it identical}. This equation is
\bea 
\del_\sigma \theta \del_t \rho - \del_t \theta \del_\sigma \rho 
- \del_\sigma h = 0.  
\label{threeseven} 
\eea
It is not hard to understand the reason for this. The action
(\ref{threesix}) is invariant under t-dependent
$\sigma$-reparametizations, $\sigma \rightarrow \sigma'(t, \sigma)$.
This gauge invariance of $S$ arises because the motion of the fluid
along the boundary is unphysical, which is due to the
indistinguishability of the underlying fermionic degrees of
freedom. As a consequence of this gauge symmetry, $\rho(t, \sigma)$
and $\theta(t, \sigma)$ provide a redundant description of the
dynamics of the boundary. The physical description of the dynamics
requires only ``half'' the number of variables. The single equation of
motion (\ref{threeseven}) describes the dynamics of this physical,
gauge-invariant degree of freedom.

For a generic potential (\ref{threeseven}) is a complicated non-linear
partial differential equation, not easy to solve. However, for the
harmonic oscillator potential it simplifies dramatically. In this case
$h=\rho$, so the equation becomes
\bea 
\del_\sigma \theta \del_t \rho - (1 + \del_t \theta)
\del_\sigma \rho = 0.  
\label{threeeight} 
\eea 
A general solution to this equation can be written down
immediately. At points on the boundary where neither $\del_\sigma
\rho$ nor $\del_\sigma \theta$ vanishes, the solution is
\bea 
\rho(t, \sigma) = f(t+\theta(t, \sigma)).
\label{threenine}
\eea
Here $f$ is any arbitrary periodic function (in $\theta \rightarrow
\theta + 2\pi$), subject only to the fixed area constraint $\int
d\sigma~\del_\sigma \theta \rho=2\pi \rho_0$. 

The above solution is not valid at the points $\sigma_i$ where either
$\del_\sigma \rho$ or $\del_\sigma \theta$ vanishes. At these points
(\ref{threenine}) requires {\it both} $\del_\sigma \rho$ and
$\del_\sigma \theta$ to vanish together, which is not possible
\footnote{There is a possibility of both $\del_\sigma \rho$ and
$\del_\sigma \theta$ vanishing for self-intersecting curves at the
point of self-intersection. However, such boundary profiles do not
seem meaningful for regions filled with fermi fluid. Moreover, a
self-intersection of the boundary is a pinching of the filled region,
which cannot be meaningfully described in the classical limit.}. The
equation of motion is, however, satisfied everywhere. Therefore, at
the points where $\del_\sigma \rho$ vanishes, we must have $\del_t
\rho=0$. Similarly, at the points where $\del_\sigma \theta$ vanishes,
we must have $\del_t \theta=-1$. To proceed further and completely
solve the classical dynamics of the equation (\ref{threeeight}), we
need to suitably gauge-fix its symmetry under t-dependent
$\sigma$-reparametizations. This is what we will do now.

\subsection{Gauge-fixing boundary reparametrizations}

We will continue our discussion with the specific example of the
harmonic oscillator potential. Generalization to other potentials is
easy and will be mentioned at the end. We start by choosing a gauge in
which $\rho$ is a fixed t-independent function of $\sigma$. This
choice is possible, except at points where $\del_\sigma \rho$
vanishes, since $\rho$ is gauge-invariant at these points. However, by
equation of motion (\ref{threeeight}), $\rho$ is still t-independent
at these points. Thus, in this gauge it is consistent with the
equation of motion to choose $\rho(t, \sigma)=\bar\rho(\sigma)$
everywhere. From (\ref{threeeight}) it follows that we must have
$\theta(t, \sigma)=-t+\bar\theta(\sigma)$, except possibly at the
points where $\del_\sigma \rho$ vanishes. But by a gauge choice we can
adjust $\theta$ to this solution at these points as well
\footnote{Since $\rho$ is gauge-invariant at the points where
$\del_\sigma \rho$ vanishes, we use the freedom of
gauge-transformations to adjust $\theta$ at these points.}. Hence, in
this gauge, the complete gauge-fixed solution of (\ref{threeeight}) is
\footnote{This solution is consistent with
(\ref{threenine}) for values of $\sigma$ for which the latter holds.}
\bea
\rho(t, \sigma) = \bar\rho(\sigma), \quad 
\theta(t, \sigma) = -t+\bar\theta(\sigma).
\label{threeninea}
\eea 
We see that each point on the boundary of the filled region simply
rotates in a circle around the origin in the phase plane. This is
precisely what happens in the fermi picture in the classical limit -
in the harmonic oscillator potential the particles simply rotate in
circles whose radii are determined by their energies.

It is easy to generalize the above argument to other potentials. Let
us rewrite the equation of motion (\ref{threeseven}) for a generic
potential as follows:
\bea 
\del_\sigma \theta (\del_t \rho - \del_\theta h) - 
\del_\sigma \rho (\del_t \theta + \del_\rho h) = 0. 
\label{threenineb} 
\eea 
As above, we gauge-fix $\rho$ such that its t-dependence is determined
by the equation $\del_t \rho - \del_\theta h=0$. By (\ref{threenineb})
this equation continues to be satisfied even at points where
$\del_\sigma \rho$ vanishes and $\rho$ is gauge-invariant. Then, from
(\ref{threenineb}) we get $\del_t \theta + \del_\rho h=0$ and by
gauge-fixing $\theta$ we can enforce this equation even at points
where $\del_\sigma \rho$ vanishes. The result is that in this gauge,
physical dynamics of the boundary is obtained by solving the equations
\bea
\del_t \rho = \del_\theta h, \quad 
\del_t \theta = -\del_\rho h.
\label{threeninec}
\eea
These are precisely Hamilton's equations for a particle with the
hamiltonian $h$. So the points on the boundary of the filled region
follow the trajectories described by the solutions to these
equations. This is exactly as expected from fermions moving in the
given potential in the classical limit.

We end this discussion by explaining the physical meaning of the above
gauge-fixing procedure. At generic points on the boundary of a filled
region, the motion of the fermi fluid can be arranged to be purely
angular or purely radial by adding an arbitrary motion of the fluid
along the boundary. The symmetry under t-dependent
$\sigma$-reparametizations allows us to do this. At the points
$\sigma_i$ where $\del_\sigma \rho$ ($\del_\sigma \theta$) vanishes,
however, the tangent to the fluid boundary is purely angular (radial),
physical motion is purely radial (angular) and only angular (radial)
motion can be changed by gauge changes. We use this latter freedom to
adjust, consistent with equation of motion, $\theta(t, \sigma)$
($\rho(t, \sigma)$) to be a smooth function of $t$ across the points
$\sigma_i$.

\subsection{Conserved charges and area-preserving diffeomorphisms}

The conserved charges that we found in (\ref{twoten}) in the exact
bosonized theory have a classical limit. In fact, the classical 
expressions for the charges can be written entirely in terms of the 
boundary variables which appear in the action (\ref{threesix}). We get
\bea
Q^{\rm cl}_n=\frac{1}{2\pi \hbar}\int^{2\pi}_0 d\sigma~\del_\sigma 
\theta~\tilde h_n, \quad n=0, 1, 2, \cdots,
\label{threeten}
\eea
where $\del_\rho \tilde h_n=h^n$. One can check directly using the
classical equation of motion (\ref{threeseven}) that these charges are
conserved. Note that, unlike the charges in (\ref{twoten}), here the
number of independently conserved charges is infinite. This is
consistent with the fact that in the classical limit the number of
fermions goes to infinity.

In the case of harmonic oscillator potential, these charges take an
especially simple form:
\bea
Q^{\rm cl}_n=\frac{1}{2\pi \hbar}\int^{2\pi}_0 d\sigma~\del_\sigma 
\theta~{\rho^{n+1} \over (n+1)}, \quad n=0, 1, 2, \cdots,
\label{threeeleven}
\eea
In addition, in this case one can give an explicit expression for
the other charges of the group of area-preserving diffeomorphisms:
\bea
W^{\rm cl}_{nm}=\frac{1}{2\pi \hbar}\int^{2\pi}_0 d\sigma~\del_\sigma 
\theta~\rho^{\frac{n+m}{2}+1}~
e^{i(n-m)\theta}, \quad n, m=0, 1, 2, \cdots.
\label{threetwelve}
\eea
Using the equation of motion (\ref{threeeight}) it can be checked that
these satisfy the equation 
\bea
\del_t W^{\rm cl}_{nm}=i(n-m) W^{\rm cl}_{nm}.
\label{threethirteen}
\eea
The ``diagonal'' charges, $W^{\rm cl}_{nn}$, are conserved. They are, in
fact, essentially the charges $Q^{\rm cl}_n$.

\section{Quantization of the collective theory}

In this section we will discuss the problem of quantization of the
collective theory. Throughout this section we will limit our
discussion to the harmonic oscillator potential. This is because in
this case quantization can be done exactly. One can, of course,
develop a perturbation expansion in small $\hbar$ for more general
potentials, but we will not pursue that here.

Our staring point is the classical action (\ref{threesix}). Canonical
quantization requires that we first fix a gauge for the symmetry under
boundary reparametrizations that this action possesses. The gauge that
we will fix is slightly different from the one we discussed in the
previous section. Here we will start by gauge-fixing $\theta$. A
convenient choice is to set $\theta(t, \sigma)=-t+\bar\theta(\sigma)$.
By arguments similar to those given in the previous section, we again
arrive at the gauge-fixed solution (\ref{threeninea}).

The choice of the precise functional form of $\bar\theta(\sigma)$
depends on whether $\del_\sigma \bar\theta(\sigma)$ vanishes anywhere
or not. Geometrically, the tangent to the boundary at such a point is
in the radial direction. Physically, at these points $\theta$ is
gauge-invariant and the gauge is fixed on $\rho$, to arrive at the
gauge-fixed solution (\ref{threeninea}). The simplest case is that of
a boundary which has no such points. More generally, however, there
may be several points on the boundary at which the tangent is radially
directed. Clearly, the number of such points has to be even because
the boundary is a closed curve, unless $\del^2_\sigma
\bar\theta(\sigma)$ also vanishes at some point, i.e. it is a point of
inflexion. Additionally, there may be points where still higher
derivatives of $\bar\theta(\sigma)$ also vanish. In the following, we
will separately discuss quantization for the two cases: (i) tangent
not radially directed at any point on the boundary and (ii) tangent
radially directed at one or more points, some of which may also have
vanishing higher derivatives of $\bar\theta(\sigma)$.

In the following, for simplicity we will restrict our detailed
discussion to a single fluid droplet, assuming further that the
droplet is centered around the origin in phase plane. At the end of
this section, we will comment on extension of this discussion to more
general fluid configurations of the type shown in Fig.1(b).

\subsection{Boundary profiles with tangent nowhere radially directed}

This is the simplest case. A representative example of this class of
boundary profiles is shown in Fig.1(a). By a gauge choice, consistent
with equation of motion, in this case we may set
\bea 
\theta(t, \sigma)=-t + \bar\theta(\sigma), \quad 
\bar\theta(\sigma)=\sigma.
\label{fourone}
\eea 
The canonical equal-time commutation relation (actually the
Dirac bracket) for $\phi$ that follows from the action
(\ref{threesix}) is
\bea
[\del_\sigma\phi(t, \sigma), \del_{\sigma'}\phi(t, \sigma')]=
-2i\pi \hbar^2 \del_\sigma \delta(\sigma-\sigma').
\label{fouronea}
\eea
Using (\ref{threefive}) and (\ref{fourone}), the commutation relation 
for $\rho$ follows:
\bea
[\rho(t, \sigma), \rho(t, \sigma')]=
-2i\pi \hbar^2 \del_\sigma \delta(\sigma-\sigma').
\label{fourtwo}
\eea
In the above gauge, $\rho$ is time-independent. In terms of modes,
\bea
\rho(t, \sigma) = \bar\rho(\sigma) = \rho_0 + \hbar
\sum_{m=1}^\infty \biggl(\alpha_m~e^{im\sigma}+
\alpha_m^\dagger~e^{-im\sigma}\biggr).
\label{fourthree}
\eea 
The constant term is fixed to be $\rho_0$ because of the fixed area
constraint $\int d\sigma~\del_\sigma \theta \rho=2\pi \rho_0$. The
physical degrees of freedom are the complex modes $\alpha_m~(m=1, 2,
\cdots)$. Because of (\ref{fourtwo}) they satisfy the harmonic
oscillator commutation relations
\bea
[\alpha_m, \alpha_n^\dagger]=m\delta_{mn}.
\label{fourfour}
\eea

The hamiltonian is given by
\bea
H=\frac{1}{4\pi\hbar}\int d\sigma~\del_\sigma \theta~{\bar\rho}^2.
\label{fourfive}
\eea
In terms of the modes of $\bar\rho$ this reads
\bea
H=\hbar \sum_{m=1}^\infty \alpha_m^\dagger \alpha_m +
\frac{\rho_0^2}{2\hbar} + ``{\rm zero-point~energy}''.
\label{foursix}
\eea 
The second term in the hamiltonian is the ground state energy. In the
fermionic picture, this is the energy of the fermi ground state. The
last term is an infinite ``zero-point energy''.

The first term in the hamiltonian gives excitation energies. The
excited states are constructed from an infinite number of decoupled
oscillators with frequencies $1, 2, \cdots$, just like in free string
theory. For low energies, these states are in one-to-one
correspondence with the spectrum of the fermion theory. However, it is
easy to see that this correspondence breaks down at high enough
energies, if the number of fermions $N$ is finite, though it may be
very large. In fact, the partition function of the fermion theory
agrees with that of the collective theory only if we cut-off the
oscillator frequency of the collective field
\footnote{I would like to thank S. Minwalla for pointing this out to
me. The calculation of the partition function of $N$ fermions in a
harmonic oscillator potential can be equivalently done as the
calculation of the partition function of a matrix valued harmonic
oscillator, gauged under $U(N)$. (In fact, the $1/2$-BPS sector of
${\cal N}=4$ superYang-Mills is actually a U$(N)$ one-matrix quantum
mechanical system with a harmonic oscillator potential \cite{CJR,DB}.)
The latter calculation must take into account only the gauge-invariant
states and the partition function over these is given by
$Z_N(\beta)=\Pi_{n=1}^N (1-e^{-\beta n})^{-1}$. Comparing with the
well-known partition function of the bosonic system described by the
Hamiltonian (\ref{foursix}), namely $Z(\beta)=\Pi_{n=1}^\infty
(1-e^{-\beta n})^{-1}$, we see that agreement requires the cut-off.}
at $N$. This cut-off has to be imposed by hand; it is not a part of
the standard quantization of the collective theory. In contrast, as we
have argued earlier, in the noncommutative (i.e. exact) formulation of
the bosonized theory, the spectrum exactly matches with the fermi
theory, for any number of fermions (including one!) without the need
for any cut-off.

Another interesting quantity to compute is the Wigner density $u(\rho,
\theta, t)$ in some state. It is possible to write down a manifestly
gauge-invariant classical expression for $u(\rho, \theta, t)$ in terms
of the functions $\rho(\sigma, t)$ and $\theta(\sigma, t)$ which
characterize the boundary. We have,
\bea
u(\rho, \theta, t) = \int d\sigma~\del_\sigma \theta(\sigma, t)~
\Theta(\rho(\sigma, t) - \rho)~\delta(\theta(\sigma, t)-\theta).
\label{fourseven}
\eea 
Here $\Theta$ is the familiar step-function. One can easily verify
from this expression that the equation of motion for $u$,
$(\del_t-\del_\theta)u=0$, implies the equation of motion
(\ref{threeeight}) for the boundary. To see that this expression
satisfies the constraint $u^2=u$, we substitute in it the gauge-fixed
solution (\ref{threeninea}) to the equation of motion. Then, the
$\sigma$ integral in (\ref{fourseven}) can be explicitly done. We get,
\bea
u(\rho, \theta, t) = \Theta(\bar\rho(t+\theta) - \rho),
\label{foureight}
\eea
which has the desired form.

Upon quantization, the boundary fluctuates, so in the quantum
theory a reasonable definition would be in terms of averages in a
given state. We define
\bea 
u(\rho, \theta, t) = <\Theta(\bar\rho(t+\theta) - \rho)>.
\label{fournine}
\eea
The quantum average can be worked out exactly in any state since we
are dealing with a free theory. Here we will restrict ourselves to the
simplest case of the ground state, $|0>$. In this case $u \equiv
u(\rho)$ will turn out to be a function of $\rho$ only. Taking a
derivative of (\ref{fournine}) with respect to $\rho$, using the
fourier representation of the resulting $\delta$-function, and doing
the average gives the result
\bea 
\del_\rho u(\rho) = -\frac{1}{\sqrt{2\pi c}}
e^{-(\rho-\rho_0)^2/2c}, 
\label{fourten}
\eea
where 
\bea
c = <0|(\bar\rho(t+\theta)-\rho_0)^2|0> = \hbar^2 \sum^\infty_{m=1} m
\label{foureleven}
\eea
To get a finite value for $c$, one needs to impose a cut-off on the 
frequency sum by hand. Then, using the boundary condition that $u(\rho)$ 
vanishes at infinity, we can integrate (\ref{fourten}) to get
\bea
u(\rho) = \frac{1}{\sqrt{2\pi c}} \int_\rho^\infty dx~
e^{-(x-\rho_0)^2/2c}.
\label{fourtwelve}
\eea
As a consistency condition, the density obtained above must satisfy
the fixed area constraint $\int d\rho~u(\rho) = \rho_0$. This
constraint \footnote{One might think that this constraint should be
automatically satisfied since, at least formally, one has
$\int_0^\infty d\rho~u=\int_0^\infty
d\rho~<0|\Theta(\bar\rho(t+\theta)-\rho)|0>=
<0|\bar\rho(t+\theta)|0>=\rho_0$.  However, it turns out this formal
argument does not work since large negative fluctuations can destroy
the positivity of $\bar\rho(\sigma)$. Hence the need to impose this
constraint explicitly.} translates into the condition
\bea
\rho_0/\sqrt{2c} = \frac{1}{2} \int_{\rho_0/\sqrt{2c}}^
\infty d\rho~(1-{\rm erf}(\rho)),
\label{fourthirteen}
\eea 
where ${\rm erf}(\rho)$ is the error function. It turns out that
this condition is always satisfied if $N_0 \leq \rho_0/\hbar$, where
$N_0$ is a cut-off (assumed large) on the sum in
(\ref{foureleven}). Since $\rho_0/\hbar$ is finite, though it may be
large for small $\hbar$, we once again see the need to impose a
cut-off on the oscillator frequencies.

The classical density function for the ground state is the
step-function $\Theta(\rho_0-\rho)$. Quantum corrections work in the
right direction and soften the fall-off to an exponential. However,
for finite $N$ the detailed functional form of the density in
(\ref{fourtwelve}) does not match with that of the exact answer for
the fermi vacuum given in (\ref{twosixteen}). For example, the
integrand in the latter case shows rapid oscillations with a
wavelength which decreases as $N^{-1/2}$ for large $N$. In contrast,
the collective theory answer for the integrand in (\ref{fourtwelve})
is a simple gaussian which has no such feature. It is important to
emphasize here that (\ref{fourtwelve}) is the exact answer for the
density in the ground state in the collective theory. There are no
corrections. We conclude that the collective quantum theory
calculation does not agree with the exact answer for finite $N$.

\subsection{Boundary profiles with radially directed tangents}

Fig.2(a) shows an example of a boundary profile which has two such
points, none of which is a point of inflexion. Fig.2(b) shows an example
of a boundary profile which has just a point of inflexion.  
\begin{figure}[ht]
\begin{tabular}{ll}
{\hspace{1.0cm}
    \epsfxsize=4.5cm
   \epsfysize=6.0cm
   \epsffile{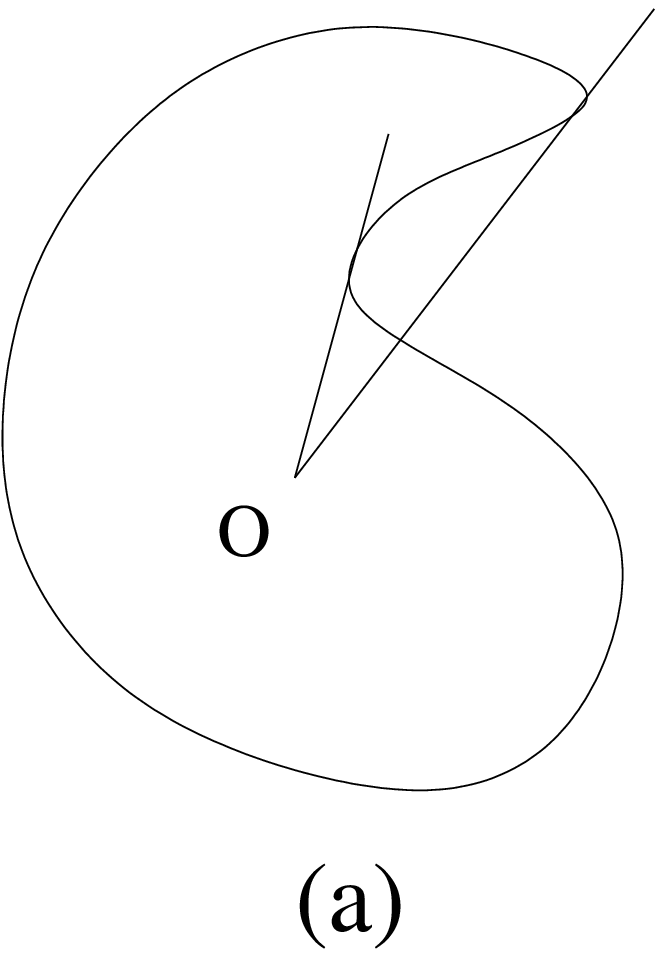}
}
&
{\hspace{3.0cm}
    \epsfxsize=4.0cm
   \epsfysize=6.0cm
   \epsffile{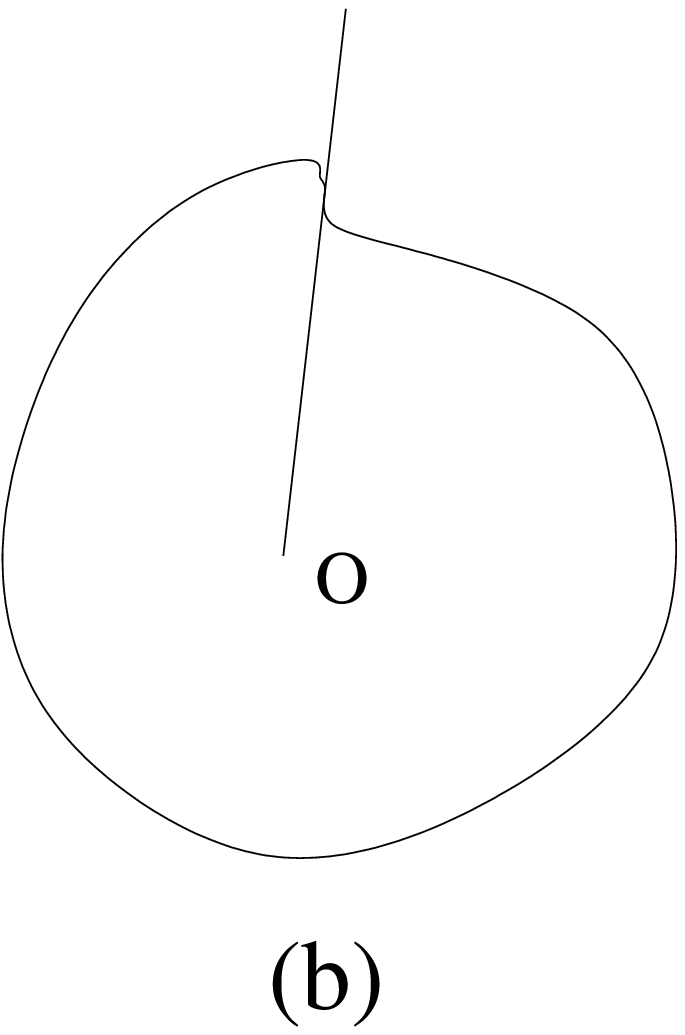}
 }
\end{tabular}
\caption{\sl 
(a) A fluid boundary profile with radially directed tangents at two
points. (b) A fluid boundary profile with a radially directed tangent
at one point which is also a point of inflexion. Both droplets are
centered at the origin O of the phase plane.}
\label{fig2ab.fig}
\end{figure}
As we have discussed, a gauge cannot be completely fixed on $\theta$
for such boundaries; gauge-fixing has to be partly done on $\rho$. As
a consequence of this, $\rho$ does not provide a complete description
of the gauge-invariant degrees of freedom for such boundaries. In this
case, a more convenient set of variables is provided by the
gauge-invariant quantities $\alpha_m$ which are defined below:
\bea 
\alpha_m
\equiv \frac{1}{2\pi\hbar}\int d\sigma~\del_\sigma \phi(\sigma, t)~
e^{-im(t+\theta(\sigma, t))}, \quad m=0, \pm 1, \pm 2, \cdots.
\label{fourfifteen}
\eea
They are manifestly invariant under time-dependent boundary
reparametrizations. Moreover, using (\ref{threefive}) and the
gauge-fixed solution to the equations of motion, (\ref{threeninea}),
we get
\bea
\alpha_m = \frac{1}{2\pi\hbar}\int d\sigma~\del_\sigma\bar\theta(\sigma)~ 
\bar\rho(\sigma)~e^{-im\bar\theta(\sigma)}.
\label{foursixteen}
\eea
For $\bar\theta(\sigma)=\sigma$, these are identical to the modes used
in the previous subsection (with $\hbar\alpha_0=\rho_0$ and
$\alpha_{-m}=\alpha_m^\dagger$) for boundary profiles without any
radially directed tangents. For the more general boundary profiles
under discussion here, there is a subtlety in inverting
(\ref{foursixteen}) to express $\bar\rho$ in terms of these modes. We
have from (\ref{foursixteen})
\bea
\hbar \sum_m\alpha_m~e^{im\bar\theta(\sigma)} = 
\int d\sigma'~\del_{\sigma'}\bar\theta(\sigma')~ 
\bar\rho(\sigma')~\delta(\bar\theta(\sigma')-\bar\theta(\sigma)).
\label{foureighteen}
\eea
Let us denote the location of zeroes of $\del_\sigma \bar\theta$ by
$\sigma_i$, where $i=1, 2, \cdots$. For $\sigma \neq \sigma_i$, we
get
\bea
\hbar \sum_m\alpha_m~e^{im\bar\theta(\sigma)} = 
{\rm sign}(\del_\sigma \bar\theta(\sigma))~\bar\rho(\sigma), \quad 
\sigma \neq \sigma_i.
\label{fourninteen}
\eea
However, for $\sigma=\sigma_i$ the right hand side vanishes, leading
to the constraint
\bea
\sum_m\alpha_m~e^{im\bar\theta(\sigma_i)} = 0,
\label{fourtwenty}
\eea
one for each point $\sigma_i$. Equations (\ref{fourninteen}) and
(\ref{fourtwenty}) express one set of gauge-invariant variables,
namely $\{\{\bar\rho(\sigma),~\sigma \neq \sigma_i\},
\{\bar\theta(\sigma_i)\}\}$
\footnote{The complimentary set, namely $\{\{\bar\rho(\sigma_i)\},
\{\bar\theta(\sigma),~\sigma \neq \sigma_i\}\}$, has been gauged away.}
in terms of another, namely $\{\alpha_m\}$. This latter set clearly
provides a more convenient starting point for a gauge-invariant
description of the quantum dynamics of generic boundary profiles.

As in the previous subsection, quantization begins with the canonical
commutation relation (\ref{fouronea}), which is valid for generic
boundary profiles. From this we deduce \footnote{One needs to use the
relation $\frac{1}{2\pi}\int d\sigma~\del_\sigma\bar\theta(\sigma)~
e^{i(m-n)\bar\theta(\sigma)}=\delta_{mn}$.} the standard harmonic
oscillator commutation relations (\ref{fourfour}) for $\alpha_m$.
Moreover, as before the Hamiltonian is given by
(\ref{fourfive}). Since $\bar\rho(\sigma_i)$ does not contribute to
the right hand side, we may use (\ref{fourninteen}) in it. This gives
precisely the expression (\ref{foursix}) for the Hamiltonian in terms
of the modes. We see that the spectrum remains unchanged, and, as
before, for finite $N$ it does not agree with the spectrum in the
fermionic theory.

Analogous to the computation following (\ref{fourseven}), here also
one may wish to compute the Wigner density $u(\rho, \theta, t)$ in
some state. First note that (\ref{fourseven}) correctly reproduces the
classical density for a generic boundary profile. Using the
gauge-fixed solution (\ref{threeninea}) to the equation of motion, we
get
\bea
u(\rho, \theta, t) = \int d\sigma~\del_\sigma \bar\theta(\sigma)~
\Theta(\bar\rho(\sigma) - \rho)~\delta(\bar\theta(\sigma)-t-\theta).
\label{fourtwentyone}
\eea 
Let $\sigma_k(t+\theta)$, $k=1, 2, \cdots$ be the points at which the
$\delta$-function clicks, i.e. $\bar\theta(\sigma_k)=t+\theta$. Then,
we get
\bea
u(\rho, \theta, t) = \sum_{k \neq i} {\rm sign}(\del_{\sigma_k} 
\bar\theta(\sigma_k))~\Theta(\bar\rho(\sigma_k) - \rho).
\label{fourtwentytwo}
\eea
The sum excludes points $\sigma_k=\sigma_i$ at which $\del_\sigma
\bar\theta$ vanishes. It is easy to see that the right hand side
precisely equals one in the filled region and is zero otherwise. The
exclusion of zeroes of $\del_\sigma \bar\theta$ from the sum as well
as the presence of the ``sign'' function in this formula is crucial to
reproduce the correct answer.

In the quantum case, we replace the step-function on the right hand
side of (\ref{fourtwentytwo}) by the average in some state. For the
ground state, the average is independent of $\sigma_k$ and has the
value given in (\ref{fourtwelve}). The sum is also trivially done
since the number of solutions, excluding the points where $\del_\sigma
\bar\theta$ vanishes, is always odd with one more plus ``sign'' than
the ``minus'' sign. The net result for the density is $+1$ times that
in (\ref{fourtwelve}), i.e. it is identical to that answer. So as
there, we conclude that the collective quantum theory calculation does
not agree with the exact answer for finite $N$.

We end this section with the following comment. In the above we have
only considered configurations with a single droplet centered around
the origin in the phase plane. For more general configurations
consisting of several disconnected droplets shown in Fig.1(b), the
density $u$ and hence the action can be written as a sum on the
different droplets. The origin of phase plane will be inside at most
one of the droplets. Figs.3(a) and (b) show examples of profiles when
the origin of the phase plane is outside the filled region.
\begin{figure}[ht]
\begin{tabular}{ll}
{\hspace{1.0cm}
    \epsfxsize=3.5cm
   \epsfysize=5.0cm
   \epsffile{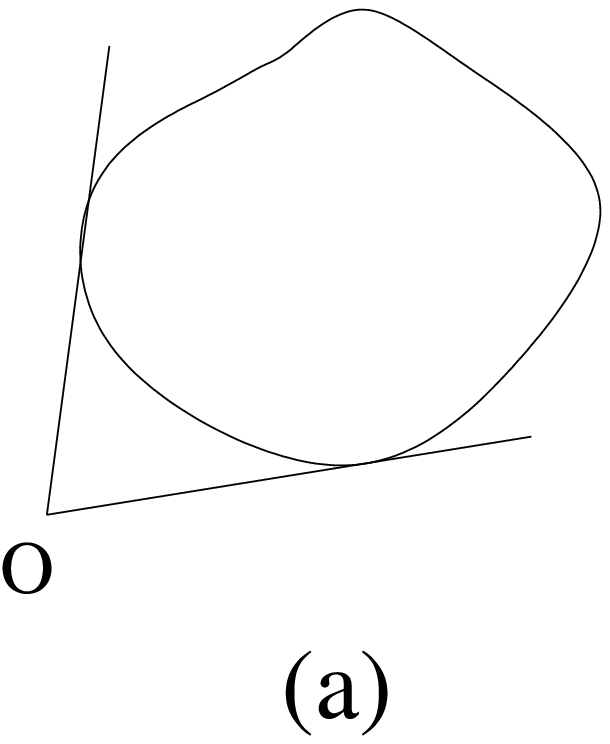}
}
&
{\hspace{3.5cm}
    \epsfxsize=5.0cm
   \epsfysize=5.0cm
   \epsffile{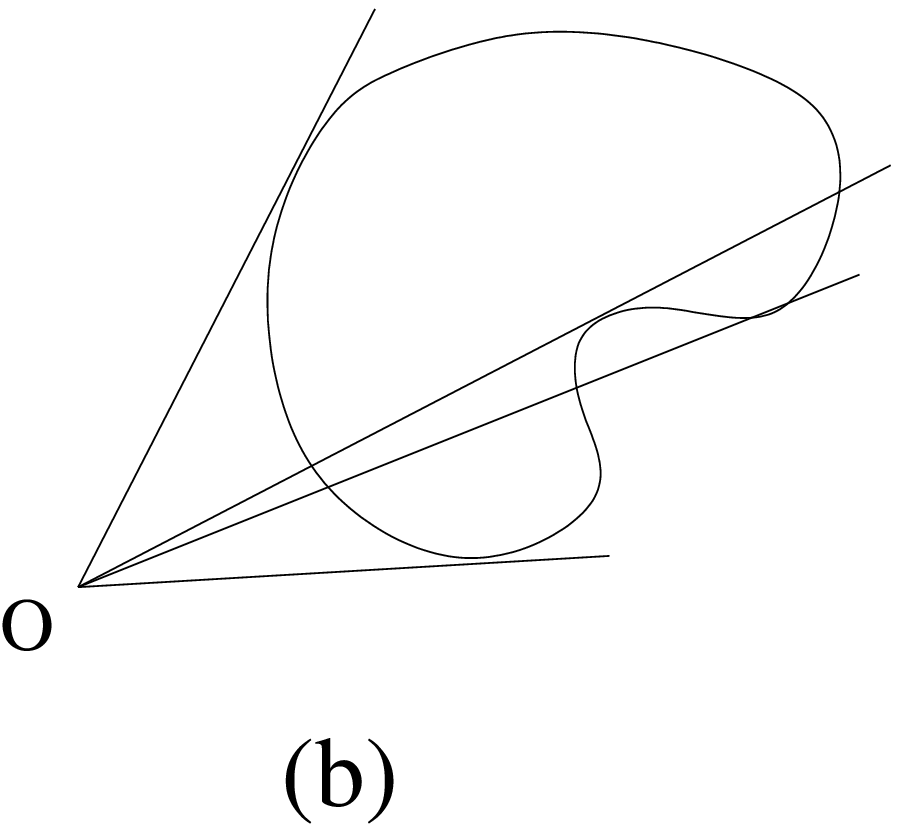}
 }
\end{tabular}
\caption{\sl 
(a) A fluid boundary profile with just the two bounding radially
directed tangents. (b) A fluid boundary profile with radially directed
tangents other than the two bounding tangents. The origin O of the
phase plane is outside both droplets.}
\label{fig3ab.fig}
\end{figure}
In such cases there are always at least two radially directed
tangents. Furthermore, $\bar\theta$ takes a maximum value which is
$<2\pi$. By suitably modifying the above discussion to incorporate
these differences, one can easily extend the present analysis to a
generic configuration of droplets.

\section{Summary and concluding remarks} 

The noncommutative theory of the Wigner phase space density developed
in \cite{DMW-classical,DMW-nonrel,DMW-path} provides an exact
bosonization of $2$-dimensional non-relativistic fermions. Utilizing
the construction of noncommutative solitons of \cite{GMS}, in this
paper we have shown that the spectrum of the bosonized theory is
identical to the spectrum of the fermion theory. Moreover, given a
Wigner density function which solves the equation of motion and
constraints of the bosonized theory, we have a precise algorithm for
building the occupied levels in the corresponding fermi state. In
contrast, we have shown that the collective quantization of
fluctuations of fermi fluid droplet boundaries neither reproduces the
spectrum nor the details of the phase space density, except perhaps
for strictly infinite number of fermions. Although we can explicitly
demonstrate this exactly only for the harmonic oscillator potential,
we believe the result is valid more generally.

One possible way out of this disagreement with the collective theory
could be that the Jacobian arising from the complicated change of
variables \cite{JS} modifies the ``classical'' action (\ref{threesix})
at finite $N$, possibly to all orders in $1/N$ (or equivalently in
$\hbar$). A systematic procedure for arriving at such a ``corrected''
action would be useful, but we have not attempted this here. Even if
one is able to find such an action, for finite number of fermions one
would still need to impose by hand a cut-off on the collective field
oscillator frequencies at $N$, which is required for the spectrum of
the collective theory to match with the spectrum of $N$ fermions. In
fact, since it is the oscillators defined in (\ref{foursixteen}) that
are the appropriate gauge-invariant variables to use for boundary
profiles of any shape, it might be useful to reformulate the
collective theory directly in terms of $N$ such oscillators, namely
the set $\{(\alpha_m,~\alpha_m^\dagger),~m=1, 2, \cdots, N\}$. Any
``corrected'' collective action in terms of this set of variables,
which reproduces the results of the fermi theory, at least
order-by-order in $\hbar$, is likely to have much of the structure of
the exact noncommutative bosonization. In fact, the latter may be a
good staring point for investigating this possibility, which we leave
for future work.

The present work has implications for probing quantum gravity in the
$1/2$-BPS sector using supergravity fluctuations \cite{GMMPR} and
D-branes \cite{GMandal}. As discussed above, exact equivalence to free
fermions via AdS/CFT requires the appearance of a noncommutative
structure on the gravity side. It would be very interesting if one
could relate this noncommutative structure to the cut-off required in
quantum gravity.

\gap3

\noindent{\bf Acknowledgments}

I would like to thank S. Minwalla for collaboration at early stages of
this work and for his comments on the manuscript. I would also like to
thank S. Wadia for discussions and comments on the manuscript.

\end{document}